\documentclass[useAMS,usenatbib]{mn2e}
\usepackage{graphicx, setspace, subfigure, latexsym, amssymb, amsmath, booktabs, wasysym, paralist}
\usepackage{multirow}
\usepackage{gensymb}

\title[The pulsar spectral index distribution]{The pulsar spectral index distribution}
\author[S. D. Bates, D. R. Lorimer and J. P. W. Verbiest]{S. D. Bates$^{1}$, 
D. R. Lorimer$^{1,2}$ and J. P. W. Verbiest$^{3}$\\
$^{1}$Department of Physics, West Virginia University, Morgantown, WV, 26506 USA\\
$^{2}$National Radio Astronomy Observatory, PO Box 2, Green Bank, WV 24944, USA\\
$^{3}$Max-Planck-Institut f\"ur Radioastronomie, Auf dem H\"ugel 69, D53121, Bonn, Germany
}
\begin{document}

\date{Accepted; \today}

\pagerange{\pageref{firstpage}--\pageref{lastpage}} \pubyear{2012}

\maketitle

\label{firstpage}
\begin{abstract}
The flux density spectra of radio pulsars are known to be steep and,
to first order, described by a power-law relationship of the form
$S_{\nu} \propto \nu^{\alpha}$, where $S_{\nu}$ is the flux density at
some frequency $\nu$ and $\alpha$ is the spectral index.  Although
measurements of $\alpha$ have been made over the years for several
hundred pulsars, a study of the intrinsic distribution of pulsar
spectra has not been carried out. From the result of pulsar surveys
carried out at three different radio frequencies, we use population
synthesis techniques and a likelihood analysis to deduce what
underlying spectral index distribution is required to replicate the
results of these surveys. 
We find
that in general the results of the surveys can be modelled by a
Gaussian distribution of spectral indices with a mean of --1.4 and
unit standard deviation. We also consider the impact of the so-called
``Gigahertz-peaked spectrum'' pulsars proposed by Kijak et al. 
The fraction of peaked spectrum sources in
the population with any significant turn-over at low frequencies
appears to be at most 10\%. We demonstrate that high-frequency ($>2$~GHz)
surveys preferentially select flatter-spectrum pulsars and the
converse is true for lower-frequency ($<1$~GHz) surveys. This implies
that any correlations between $\alpha$ and other pulsar parameters
(for example age or magnetic field) need to carefully account for
selection biases in pulsar surveys. We also expect that many known pulsars
which have been detected at high frequencies will have shallow, or positive,
spectral indices. The majority of pulsars do not have recorded flux density measurements
over a wide frequency range, making it impossible to constrain their spectral shapes. 
We also suggest that such measurements would allow an improved description
of any populations of pulsars with `non-standard' spectra.
Further refinements to this picture will soon be possible from the results
of surveys with the Green Bank Telescope and LOFAR.
\end{abstract}


\section{Introduction}

Radio pulsars are known to have steep flux-density spectra
\citep[e.g.][]{sieber1973} which approximately follow the power-law
relationship
\begin{equation}
S_\nu \propto \nu^{\alpha}
\label{eq:spectralindex}
\end{equation}
for observed flux density $S_\nu$ at frequencies $\nu > 100~\mathrm{MHz}$
where $\alpha$ is the spectral index. Although a number of
multi-frequency flux measurements of pulsars were carried out over the
years \citep{sieber1973,bf74,mm80,sab86},
the first systematic study of the spectral indices for a large number
of pulsars was carried out by Lorimer et al.~(1995) who published spectra
on 280 pulsars based on flux density measurements carried out for
up to five different radio frequencies between 0.4 and
1.6~GHz \nocite{lorimer1995} using the Lovell radio telescope.  The
resulting analysis showed that the average spectral index for the
observationally selected pulsar sample was $\alpha = -1.6 \pm
0.3$. More recent work by \citet{maron2000} derived a mean value of
$\alpha = -1.8 \pm 0.2$, where the range of frequencies was extended,
in many cases, up to 5~GHz. \citet{maron2000} also found evidence for $\sim 10\%$ of
pulsars in their sample being best fit using a double power law,
described by parameters $\alpha_1$ at lower frequencies and $\alpha_2$
at higher frequencies. Typically $\alpha_2$ is steeper, $|\alpha_2| > |\alpha_1|$,
with the spectral break at a frequency of $\sim 1 \mathrm{GHz}$.

\begin{table*}
	\begin{center}
	\caption{Observational parameters for the three pulsar surveys discussed in the text: the Parkes southern pulsar survey \citep[PKS70, ][]{mld+95}, the Parkes multi-beam pulsar survey \citep[PMPS, ][]{mlc+01} and the Parkes 6.5~GHz multi-beam pulsar survey \citep[MMB,][]{bates2010}.}
		\begin{tabular}{lccc}
		\toprule
		& PKS70 & PMPS & MMB \\
		\midrule
		%\multicolumn{2}{c}{Receiver Parameters} \\
		Number of beams  & 1 & 13 & 7 \\
		Polarizations/beam   & 2 & 2 & 2 \\
		Centre frequency (MHz) & 436 & 1352 & 6591\\
		Frequency channels  & 256 & 96 & 192 \\
		Channel width (MHz) & 0.125 & 3 & 3\\
		Gain ($\mathrm{K Jy}^{-1}$) & 0.64 & $\sim 0.7$ & 0.6 \\
		\\
		Integration time (s) & 157.3 & 2100 & 1055 \\
		Sampling interval ($\mu$s) & 300 & 250 & 125\\
		\\
		\multirow{2}{*}{Region covered} & \multirow{2}{*}{$\delta < 0\degree$} & $100\degree \leq l \leq 50\degree$ & $-60\degree\leq l \leq 30\degree$\\
		& & $|b| \leq 5\degree$ & $|b| \leq 0.25\degree$ \\
		Pulsars detected ($P>50~\mathrm{ms}$) & 279 & 1038 & 18\\
		\bottomrule
		\end{tabular}
		\label{table:survpars}
	\end{center}
\end{table*}

An interesting additional effect noticed by Maron et al. was the
high-frequency spectral turnovers observed for PSRs~B1823--13 and
B1838--04. While turnovers are frequently observed at frequencies
below $\sim100$~MHz \citep[e.g.][]{iz1981}, the presence of such
turnovers at frequencies around 1~GHz seems to be unusual. Recent
research on a subset of these pulsars (not including PSR~B1838--04) 
with so-called gigahertz-peaked spectra
\citep[hereafter GPS; see][]{kijak2011b, kijak2011a} indicates that these
sources can be fitted by
\begin{equation}
S_\nu = 10^{ax^2 + bx + c} ; \quad x = \log_{10}\nu,
\label{eq:gps}
\end{equation}
an extension of Equation~\ref{eq:spectralindex}, where $b$ is
equivalent to $\alpha$ in the case where $a=0$. 
\citet{kijak2011b} list five known GPS pulsars, with measured
values of the turnover parameter, $a$, ranging from $-7.3$ 
to $-1.2$. However, both they and, 
more recently, \citet{dembska2012}
observe that PSR~B1259$-$63 displays many different spectral types
depending upon the orbital phase when flux measurements are taken ---
including a GPS-type spectrum when near periastron. They
conclude that non-standard spectral shapes in the pulsar population
could be linked to unusual environments around pulsars.

While the observed distribution of pulsar spectra has been the subject
of much investigation \citep{lorimer1995,maron2000}, the underlying
distribution for the population is currently not understood. Knowledge
of this distribution has important implications for the pulsar
emission mechanism, and for derivation of the pulsar luminosity
function over a broad range of radio frequencies.  The latter point
has been largely overlooked so far, with population analyses
\citep[e.g.][]{lfl+06,fk06} choosing to focus on the hugely successful
20-cm Multibeam Pulsar Surveys carried out at Parkes \cite[for a
review, see][]{Lyne08}. The advantage of this approach is that it
does not require any knowledge of the radio spectrum to constrain the
population of sources visible at 20-cm. The disadvantage is that it
precludes any reliable statements being made about the population of
pulsars visible at other observing frequencies.

In this paper, we aim to use detailed Monte Carlo simulations of pulsar surveys
carried out at multiple frequencies to constrain the underlying distribution
of $\alpha$ in the pulsar population. The focus of
this work is on the normal (i.e.~non-recycled) population. Although
millisecond pulsar spectra appear to be similar to normal pulsars
(see, e.g., Kramer et al.~1998), a detailed study of this population
will be deferred to a future paper. In \S~2 we
review the surveys which define our sample. In \S~3 we
describe the relevant population modelling techniques used in this
study. In \S~4 we use our sample to place constraints on the
underlying spectral index distribution assuming it to be a simple
power-law, and discuss how our results would change were the surveys found to 
be incomplete. In \S~5 we investigate what fraction of the population
could display alternative spectral shapes.
Finally, the implications of our results are discussed in \S~6.

\begin{table}
	\begin{center}
	\caption{Parameters used to simulate the pulsar population in \S \ref{subsec:method1}.}
		\begin{tabular}{lr}
		\toprule
	     Radial distribution model & \citet{lfl+06} \\
		 Galactic latitude scale height & 330 pc \\
		 \\
		 Luminosity distribution & Log-normal \\
		 $\langle{\log\,\mathrm{L~(mJy~kpc^2)}}\rangle$ & $-1.1$\\
		 $\mathrm{std}(\log\,\mathrm{L~(mJy~kpc^2)}))$ & $0.9$\\
		 \\
		 Period distribution & Log-normal\\
		 $\langle{\log\,\mathrm{P~(ms)}}\rangle$ & $2.7$\\
		 $\mathrm{std}(\log\,\mathrm{P~(ms)})$ & $-0.34$\\
		 \\
		 Scattering model & \citet{bcc+04}\\
		 \\
		 Number of detectable pulsars & \multirow{2}{*}{1038}\\
		 in the PMPS survey \\%1201\\
		\bottomrule
		\end{tabular}
		\label{table:popparams}
	\end{center}
\end{table}

\section{Pulsar surveys}

Many surveys of the Galactic plane for pulsars have been performed, at
different radio frequencies, despite the steep spectral behaviour
discussed in the previous section. This is because of two effects
which limit sensitivity at low observing frequencies. Firstly
synchrotron radiation from free electrons in the Galactic magnetic
field causes the sky background temperature, $T_\mathrm{sky}$, to vary
with frequency as $\nu^{-2.6}$ \citep{lmop87}. At low frequencies, the
sky temperature can then dominate the system temperature and reduce
sensitivity to pulsars. Secondly, as signals from pulsars traverse the
interstellar medium, they are scattered, which causes them to be
broadened by a timescale which scales approximately as $\nu^{-3.5}$
for high-DM pulsars \citep{lkm+01,bcc+04}. Since this effect is not easily removed,
low-frequency surveys are severely hampered when searching for faint
pulsars in the Galactic plane. As a result of these issues,
pulsar surveys have to compromise between these two effects, and
the reduced flux density which is observed from pulsars at higher
frequencies. Three examples of pulsar surveys at three observing
frequencies are outlined below. Table~\ref{table:survpars} summarises
each of these surveys.

By far the most successful pulsar survey to date was the Parkes
multi-beam pulsar survey \citep[hereafter the PMPS survey, ][]{mlc+01}
which, to date, has discovered close to 800 pulsars, essentially
doubling the number previously known.  The PMPS observed a thin strip of the
Galactic plane, Galactic latitude $|b|<5\degree$, Galactic longitude
$100\degree<l<50\degree$, at a frequency of 1.4~GHz using the 64-metre
Parkes radio telescope. The Parkes southern pulsar survey
\citep[PKS70, ][]{mld+95} was a survey of the entire southern sky,
$\delta<0\degree$, at a frequency of 436~MHz. The survey discovered
101 pulsars, many of which were at high Galactic latitudes. The Parkes
6.5~GHz multi-beam pulsar survey \citep[MMB,][]{bates2010} used the
seven-beam Parkes Methanol Multi-beam receiver to survey the Galactic
plane in the region $-60\degree\leq l \leq 30\degree,
|b|\leq0.25\degree$ at a frequency of 6.5~GHz. This survey was
specifically designed to discover pulsars deep in the Galactic plane
which would otherwise be obscured by the interstellar scattering
discussed above. This survey discovered two pulsars, indicating that
there are few pulsars in the direction of the Galactic centre whose
spectra obey a shallow or flat power law.

\begin{figure*}
	\begin{center}	
		\includegraphics[width=17cm]{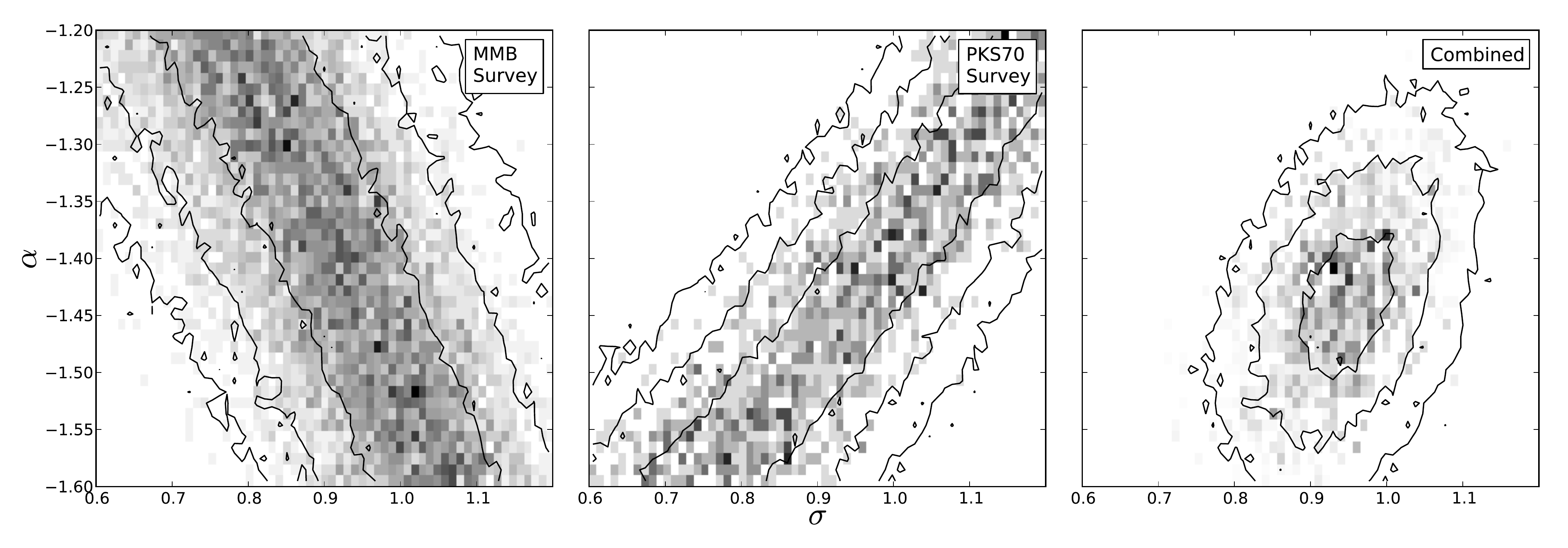}
	\end{center}
	\caption{Grey scale plots of the likelihood, ${\cal L}$ (as defined in \S~\ref{sec:analysis1}) as a function of the mean, $\alpha$, and standard deviation , $\sigma$, of the input spectral index distribution. The left and centre plots show the results for the MMB and PKS70 surveys respectively, and the right-hand plot is the product of the likelihoods in each bin. Significance contours of 1, 2 and 3 standard deviations --- calculated by Equation~\ref{eq:confidence} --- are overlaid. The total confidence levels in the right-hand panel were calculated by summing the MMB and PKS70 survey confidence levels in quadrature. }
	\label{fig:siresults}
\end{figure*}

\section{Population synthesis}

There are two main techniques which are commonly applied when
simulating the pulsar population \citep[for a recent review, see][]{lorimerheep2011}. 
The first is the ``snapshot'' approach, which
is to synthesise a pulsar population in the current epoch based
entirely upon the known statistics of the observed population
\cite[see, e.g.,][]{lfl+06}.  The second is a more
computationally-involved, fully dynamical technique, where a model
galaxy of pulsars is given initial birth positions and parameters,
then allowed to evolve using models of the pulsar spin-down and
Galactic potential \cite[see, e.g.,][]{fk06,rl10a}. The model
population can be compared to the results of previous pulsar surveys
to check its validity.

In this work, where the main focus is to explore the dependence of pulsar
spectra on survey yields, 
we will use the snapshot approach which will make no assumptions
about the birth parameters of pulsars, or their subsequent evolution.
Instead we are able to produce model populations using parameters obtained
purely from observations --- though it must be remembered that this
technique is still reliant on some models, for example the Galactic
electron distribution, for which no one model entirely describes the
observed data. Following standard practice, we adopt the model of 
Cordes \& Lazio (2002) \nocite{cl02b} in our simulations described below.

\section{Obtaining the spectral index distribution of normal pulsars}\label{sec:method1}
\subsection{Method}\label{subsec:method1}

For all our model populations, we adopted the parameters given in
Table~\ref{table:popparams}, and used the \textsc{psrpop} software
package\footnote{http://psrpop.sourceforge.net} \citep[based upon work
  done for][]{lfl+06} to generate synthetic pulsar populations which
we grew in size until we obtained a sample of 1038 model pulsars
detected by the PMPS survey. The number of detections in the MMB and
PKS70 surveys was not constrained.  The parameters were chosen to
match the studies of \citet{lfl+06} and \citet{fk06}, who found that a
log-normal luminosity distribution accurately describes the known
population. To simulate the spectral properties of pulsars, we
initially adopted a simple approach and assumed that the pulsar
spectral indices are normally distributed \citep[as in][]{lorimer1995,
  maron2000}.  Subsequent modifications to this technique are
described in \S \ref{sec:method2} and \ref{sec:method3}.  For this
simple model, we allowed the mean spectral index, $\alpha$, and
standard deviation, $\sigma$, of this distribution to vary in the
range $-2.0<\alpha<-1.0$ and $0.0<\sigma<2.0$ (each in steps of 
0.01). Each combination of
$\alpha$ and $\sigma$ was realised 500 times using different starting
seeds for the random number generators. The resulting set of 500 model
populations were analysed as described below.

\subsection{Analysis}\label{sec:analysis1}

We selected the sample of observed pulsars based on the results of the
PKS70, PMPS and MMB surveys with which we could compare the
simulations. From the ATNF Pulsar Catalogue \citep{mhth05}, there are
a total of 1197 pulsars with periods $P>50~\mathrm{ms}$ detected in
one or both of the PKS70 and PMPS surveys. From \citet{bates2010}, 18
pulsars were observed in the MMB survey, four of which were not
observed in the PKS70 or PMPS surveys. This gives a total of 1201
pulsars which have been observed in one or more of the surveys;
Table~\ref{table:survpars} gives a breakdown of the number of pulsars
by survey. The aforementioned period cut-off effectively removes the
recycled pulsars from this sample and none of the pulsars in the list
of 1201 is associated with any known globular cluster.

The \textsc{psrpop} program \textsc{survey} was used on each of the
model populations to create a list of detected pulsars for the
surveys. To measure the effectiveness of our models at reproducing
the observed survey yields, we adopted the following simple likelihood
analysis. For each bin in the $\alpha$-$\sigma$ space, two likelihood 
values, ${\cal L}_{\rm MMB}$ and ${\cal L}_{\rm PKS70}$, were computed. 
These likelihoods are defined to be
the fraction of 500 realisations for which 
the simulated survey detected the same number of pulsars as in the
catalogue, for both the MMB and PKS70 surveys. When presenting our
results below, we show the individual likelihoods, and their product
${\cal L}={\cal L}_{\rm MMB} \times {\cal L}_{\rm PKS70}$. The maximum
value of ${\cal L}$ allows us to constrain the ranges of $\alpha$ and
$\sigma$ which best match the observed survey yields.

Confidence levels, $\cal C$, were also independently calculated in 
each $\alpha$-$\sigma$ bin,
\begin{equation}
{\cal C} = \frac{\bar{n}_\mathrm{sim} - n_\mathrm{cat}}{\mathrm{stddev}(n_\mathrm{sim})}
\label{eq:confidence}
\end{equation}
where $\bar{n}_\mathrm{sim}$ is the mean number of pulsars detected in each realisation,
$n_\mathrm{cat}$ is the number of pulsars detected in the survey, according to
the pulsar catalogue, and $\mathrm{stddev}(n_\mathrm{sim})$ is the standard deviation
of $n_\mathrm{sim}$ across the 500 realisations.

\subsection{Results}

\begin{figure}
	\begin{center}	
		\includegraphics[width=8cm]{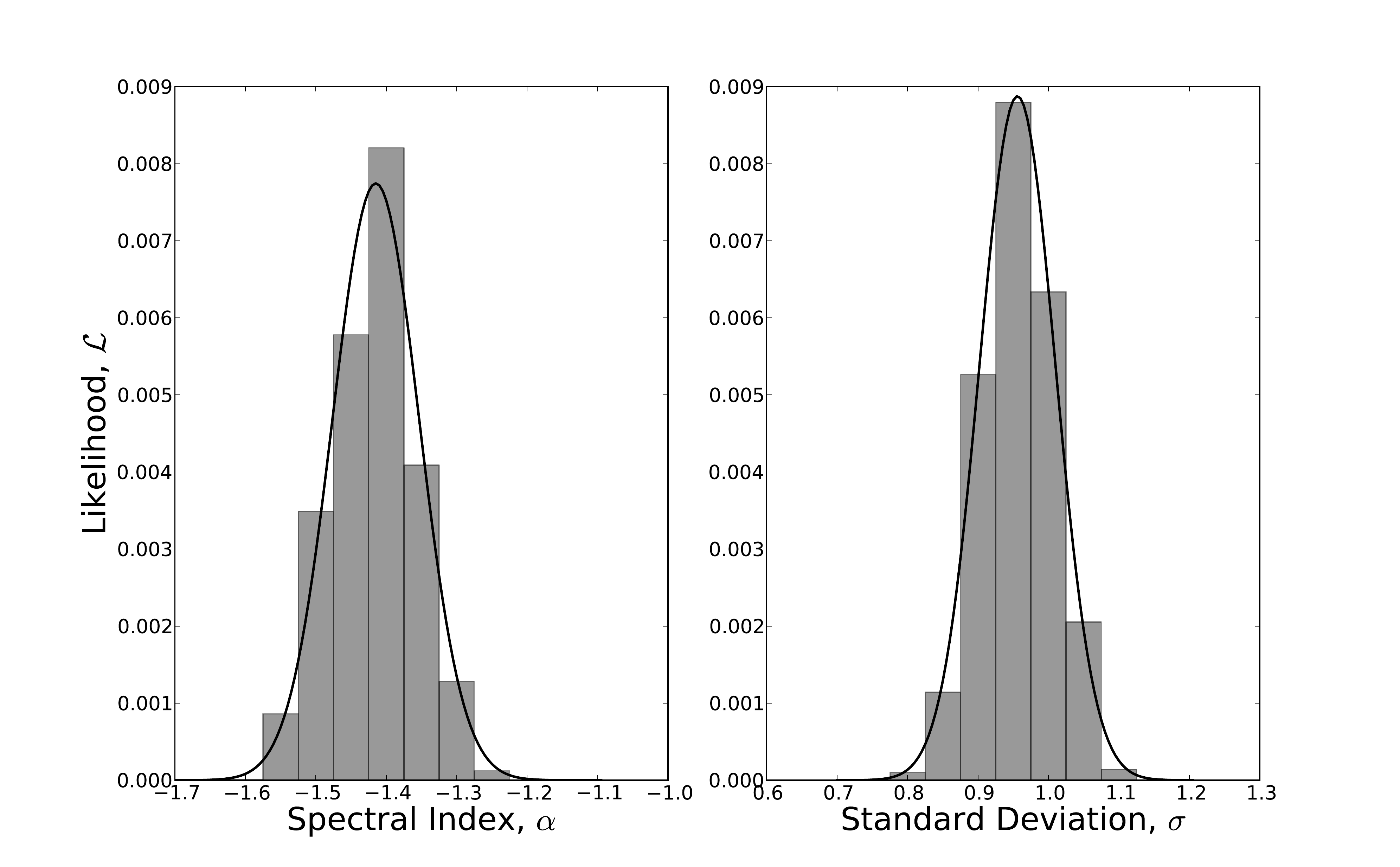}
	\end{center}
	\caption{Marginalized probability density functions obtained for each 
spectral index (left) and standard deviation (right) bin used in our simulations.
Best fits are shown with a solid line, giving the results $\bar{\alpha} = -1.41$
and $\bar{\sigma} = 0.96$.}
	\label{fig:marginalisedresults}
\end{figure}

\begin{figure}
	\begin{center}	
\includegraphics[width=8cm]{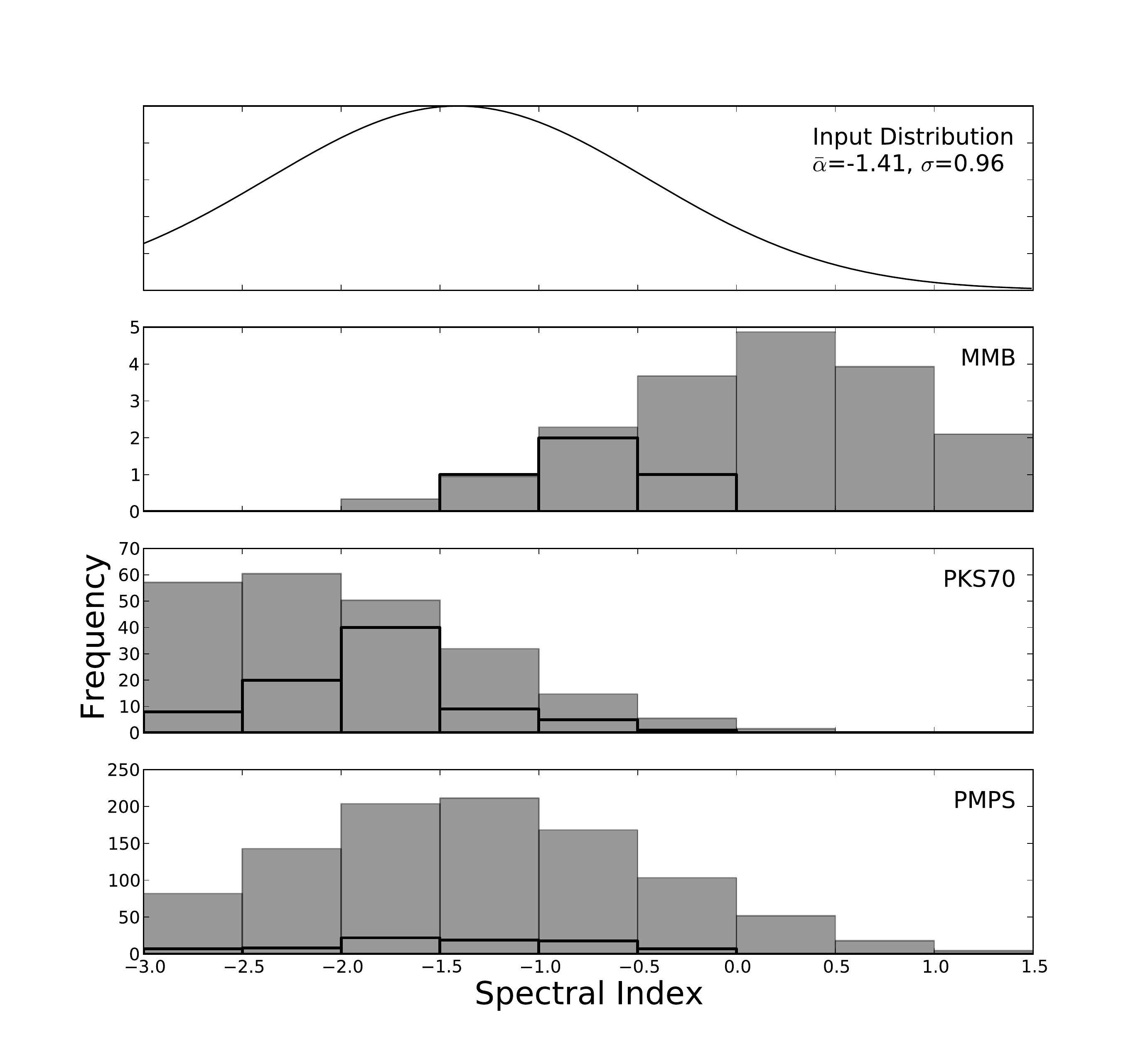}		%\includegraphics[width=8cm]{/Users/sbates/Desktop/scripts/data/finalSIdists/results/det2_-175_045.png}
	\end{center}
	\caption{Histograms of the spectral index of detected pulsars
          in our simulations (grey; averaged over 100 realisations)
          and from the pulsar catalogue (black outline) for the optimum
          parameters obtained from
          Figure~\ref{fig:marginalisedresults}. The top panel shows
          the input distribution used to generate a model pulsar
          population. Note that the spectral index values in the 
          pulsar catalogue are both incomplete, and biased towards 
          those pulsars which have been discovered in early low-frequency
          surveys.}
	\label{fig:combinehists}
\end{figure}

The results of the analysis procedure which gives the likelihoods,
${\cal L}_{\rm MMB}$ and ${\cal L}_{\rm PKS70}$, for the MMB and PKS70
surveys are shown in Figure~\ref{fig:siresults}. Note that the region of high
likelihood in the $\alpha$-$\sigma$ plane is oriented in a reasonable way;
the MMB survey only allows steep values of $\alpha$ if there is a
large width in the distribution, whereas the PKS70 survey only allows
shallow $\alpha$ values if the distribution is wide. The product of
those results, ${\cal L}$, is also shown in the right-hand panel,
which gives the overall result shown in
Figure~\ref{fig:marginalisedresults}, where the likelihoods have been
summed in the $\alpha$ and $\sigma$ directions to produce marginalized
probability density functions. Fitting a gaussian to
these distributions provides a very straightforward way to quantify
them. From these fits, we find the optimum spectral index distribution to
be $\bar{\alpha} = -1.41 \pm 0.06$ and $\bar{\sigma} = 0.96 \pm 0.05$.  To
illustrate how the models fit the data in the catalogue,
Figure~\ref{fig:combinehists} shows average spectral index histograms
for a further 100 realisations performed with the values of $\bar{\alpha}$
and $\bar{\sigma}$ obtained above.

In the early stages of this work, we had initially hoped to use the
observed spectral index distributions obtained from the pulsar
catalogue (the majority of these values are taken from
\citet{lorimer1995}) as part of the analysis, however, the
completeness of these samples was found to be too low to produce
reliable results. Only 4/18 pulsars in the MMB survey have measured
spectral indices,
84/279 pulsars in the PKS70 survey, and 83/1038 pulsars in the PMPS survey. 
As the catalogue values of $\alpha$ are so incomplete,
Fig.~\ref{fig:combinehists} serves as something of a prediction for
further measurements of the spectral indices of pulsars detected in
the MMB and PKS70 surveys --- there are a dearth of pulsars which so
far have had their spectral indices measured to be $\alpha<-2.0$ or
$\alpha>0.0$. Further observational work in this area would be
extremely valuable in refining this model, and our subsequent
modifications to it discussed below.

\subsection{Checking for biases in our results}

The number of detected pulsars for each survey, as shown in
Table~\ref{table:survpars}, was obtained from the most recent version
of the ATNF Pulsar Catalogue \citep{mhth05}.  However, these numbers
are by no means absolute --- the PMPS survey data have been
reprocessed several times \citep[e.g.\,][]{fsk+04, eatough2010,
  mickaliger2012}, each time increasing the recorded number of pulsar
detections. Furthermore, the MMB survey yielded its second pulsar
discovery only after reprocessing of the data
\citep{bates2010}. Therefore, the numbers shown in
Table~\ref{table:popparams} are lower limits; although the rate of new
detections in old data should decrease over time as successively
fainter objects are found.

It is important to quantify what impact an increase in the number of
detected pulsars in the PMPS, PKS70 and MMB surveys would have on our
results.  Firstly, to evaluate the impact of further pulsars being
discovered in the PKS70 and MMB surveys, the likelihoods ${\cal
  L}_{\rm MMB}$ and ${\cal L}_{\rm PKS70}$ were recalculated assuming
the PKS70 and MMB survey yields increased over the range of 5\% to
50\%.

In the case where the PMPS survey yield does not change (since such a
large number of pulsars have been detected in the survey data, this is
not an unreasonable approximation), the calculated values of
$\bar{\alpha}$ and $\bar{\sigma}$ do not change significantly as the
MMB survey yield increases.  As the PKS70 yield is increased,
$\bar{\alpha}$ and $\bar{\sigma}$ change by 2 and 3 standard
deviations if the yield increases by 15 and 20\%, respectively (this
corresponds to the discovery of a further 42 and 56 pulsars in the
data).  Repeating this analysis with the PMPS survey yield increased
by 5 and 10\% (a large increase on 1038 pulsars) does not change the
conclusion --- unless the three surveys are all woefully incomplete,
$\bar{\alpha}$ and $\bar{\sigma}$ do not deviate significantly from
our best-fit values.

\section{Alternative Spectral Shapes}\label{sec:spectralshapes}
\subsection{Setting limits on the size of the GPS population}\label{sec:method2}
\subsubsection{Method \& Analysis}\label{subsec:method2}

The \textsc{populate} code (part of
\textsc{psrpop}) was modified to enable the simulation of
GPS pulsars.  Additional user options were added to specify the
fraction, $F$, of pulsars which show GPS behaviour, and to
specify the value of the turnover parameter, $a$, in Equation~\ref{eq:gps}.

As the population is generated, $F$ of the pulsars are randomly
selected to be GPS sources, but are allocated a flux at 1.4~GHz and a
spectral index, $\alpha$, in the usual way. The value of $c$ in
Equation~\ref{eq:gps} may then be calculated (with $x$ defined as in
Equation~\ref{eq:gps}) as
\begin{equation}
c = \log_{10}S_{1.4} - ax^2 - \alpha x
\label{eq:gps_c}
\end{equation}
since, as stated earlier, $b$ is considered equivalent to
$\alpha$. The flux of the GPS pulsars in the simulation can then be
calculated at other frequencies using Equation~\ref{eq:gps}.% (see
%Figure~\ref{fig:gpsfluxes}).

Following the procedure used in \S~\ref{subsec:method1},
simulations were performed in order to fill out a grid of $F$ and $a$
values. To reduce the number of variables, the spectral index
mean and standard deviation were held fixed at $\alpha = -1.41,
\sigma=0.96$, which are the best values from the previous
simulations. Again, 500 realisations were performed at each grid
point, and a likelihood was calculated in the same way as in
\S~\ref{subsec:method1}.

\subsubsection{Results}

\begin{figure*}
	\begin{center}	
		\includegraphics[width=17cm]{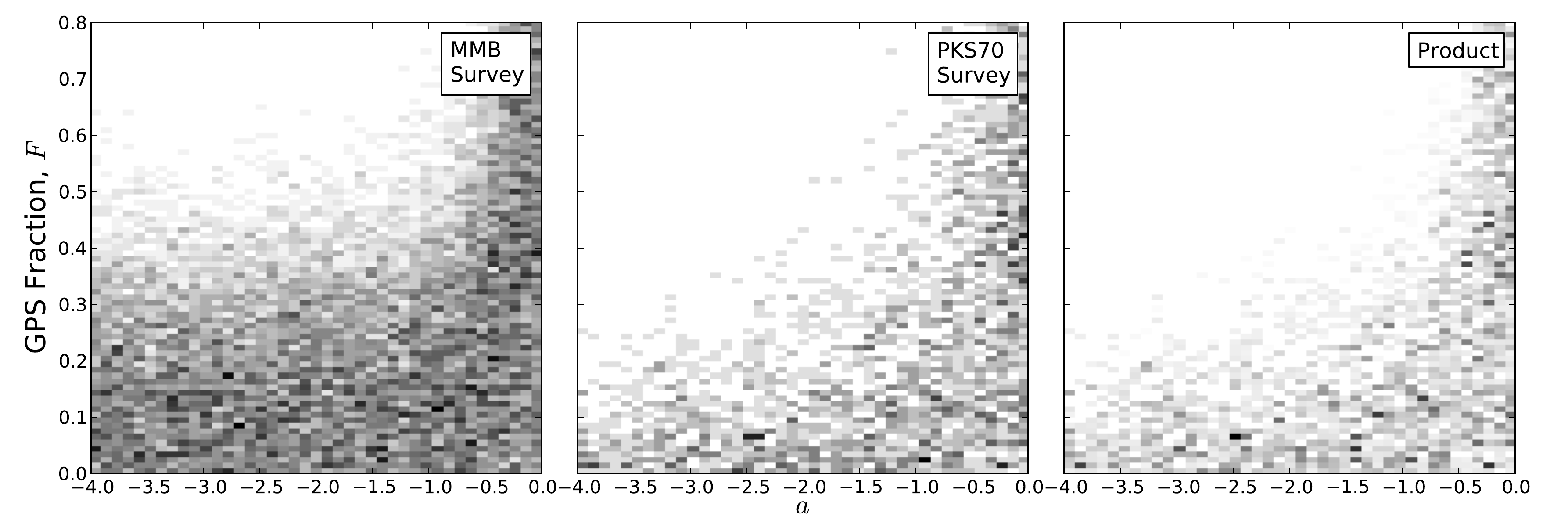}
	\end{center}
	\caption{Grey scale plots showing the likelihood (as defined in \S~\ref{sec:analysis1} as a function of the fraction of GPS pulsars in the simulation, $F$, and the turnover parameter used in Equation~\ref{eq:gps}, $a$.}
	\label{fig:gpsresults}
\end{figure*}

Figure~\ref{fig:gpsresults} shows the likelihoods calculated from the
simulations as a function of $F$ and the turnover parameter,
 $a$. Significant values for $a$,
indicating a severe turnover, are only allowed when $F$ is below
$\sim10\%$. Otherwise, values of $a$ as small as $-0.5$ are allowed in
a large fraction, $F \geq 0.40$, of the pulsar population.

\begin{figure*}
	\begin{center}	
		\includegraphics[width=11cm]{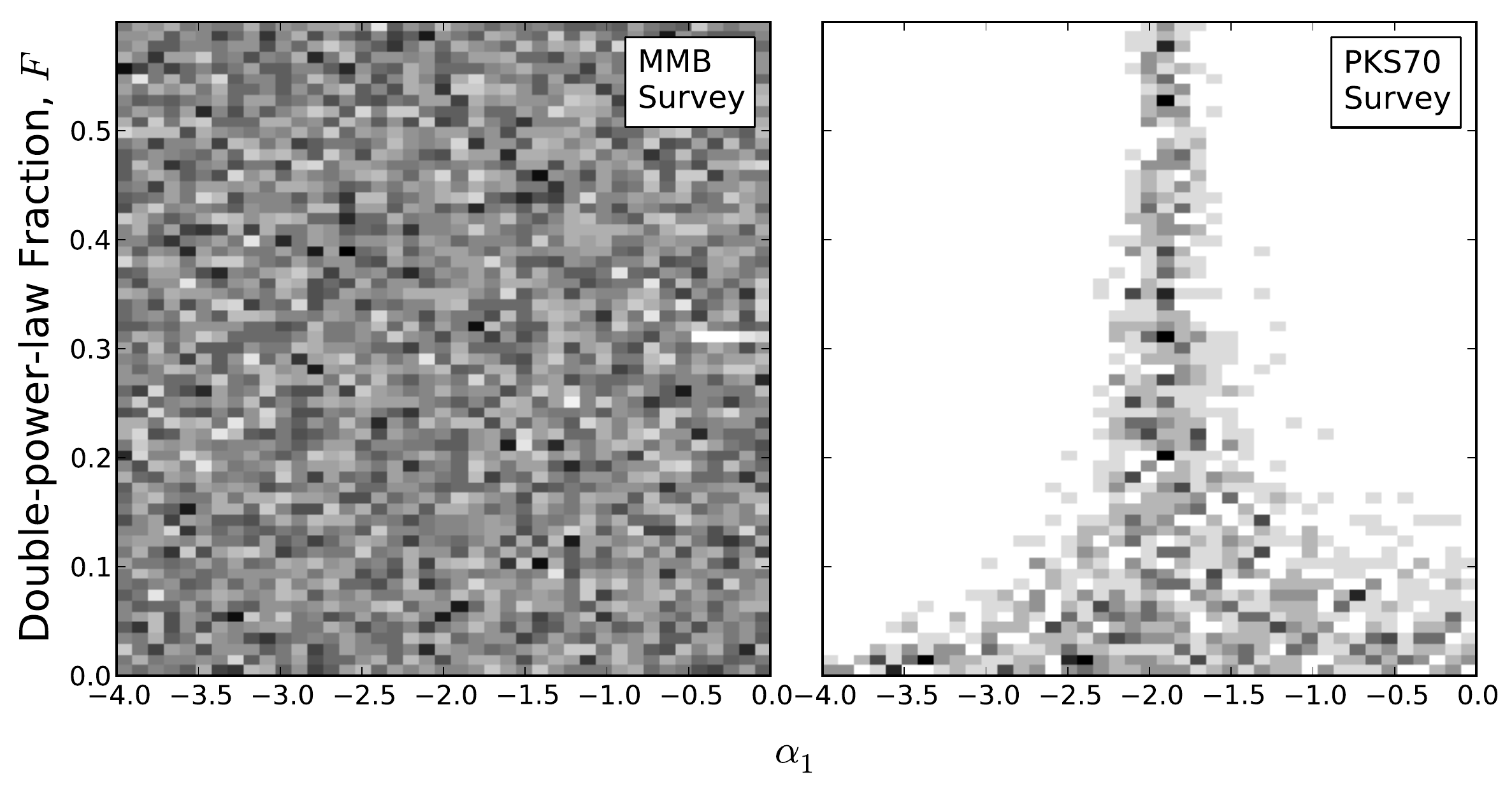}
	\end{center}
	\caption{Grey scale plots showing the likelihood (as defined in 
	\S~\ref{sec:analysis1} as a function of the fraction of double-power-law
	pulsars in the simulation, and the value of $\alpha_1$ used. In this case, 
	the product of the likelihoods is unconstrained by the MMB survey,
	and so resembles the right-hand plot.}
	\label{fig:brkresults}
\end{figure*}

\subsection{Setting limits of the size of the double-power-law population}\label{sec:method3}
\subsubsection{Method \& Analysis}\label{subsec:method3}

Further modifications were made to the \textsc{populate} code to allow
any given fraction, $F$, of pulsars to have double power-law
spectra. As the population is generated, $F$ of the pulsars are
randomly selected and given a power-law value, $\alpha_1$, to be used
at frequencies below 1.4~GHz. The flux density and ``high frequency''
$\alpha_2$ are calculated in the usual way.
As in \S~\ref{subsec:method2}, simulations were performed,
filling out a grid of $F$ and $\alpha_1$ values, holding fixed
$\alpha_2 = -1.41, \sigma=0.96$. Again, 500 realisations were
performed at each grid point, and a likelihood was calculated in the
same way based upon the number of realisations that were able to
reproduce the number of detections.

\subsubsection{Results}

The results from these simulations are shown in
Figure~\ref{fig:brkresults}. The first point of note is that the
likelihoods for the MMB survey are not affected by changing values of
$\alpha_1$, as expected. For the PKS70 survey, though, these two
parameters have a substantial impact.  We can constrain the population
of pulsars with double-power-law spectra to be below $\sim 10\%$ ---
above this fraction, the value of $\alpha_1$ is fixed at $-1.9$,
consistent with a single power law (cf.~centre panel of
Figure~\ref{fig:siresults} for $\sigma=0$). Within this 10\% of
sources, the value of $\alpha_1 $ is not well constrained, though
shallower spectral indices, $\alpha_1 > -2$ are favoured.

\subsection{Limitations and future work}
Having investigated these alternative spectral shapes for pulsars,
we are unfortunately not able to place clear constraints upon 
the size of such populations, or the values of the parameters used
to describe their spectral shapes. The only way that such work 
will be possible is with the results of on-going, and future, surveys
at a variety of observing frequencies. Two such on-going pulsar
surveys are described below, and in Table~\ref{table:simsurveypars}.

The Green Bank Northern Celestial Cap survey\footnote{For an up-to-date 
list of pulsar discoveries from this survey, see http://arcc.phys.utb.edu/gbncc.} (GBNCC) is an on-going
pulsar survey of the Northern sky using the 100-m Green Bank Radio
Telescope in West Virginia. Phase~1 of the survey has covered the
entire sky with declination $\delta \geq +38\degree$ (Stovall~et
al.~in prep.).

LOFAR is an interferometer based in Europe
which is able to observe the sky over two low-frequency
bands; one from 30-80~MHz and the other from 110-240~MHz. Simulations
for a pulsar survey using LOFAR were first performed by \citet{bws}
assuming that the survey would be performed in the higher radio band
due to pulse smearing at lower frequencies. These simulations
estimated that such a survey with LOFAR would detect $1100 \pm 100$
normal pulsars.

With the additional information provided by such large-scale surveys,
it may be possible to constrain the size of any population of pulsars
with these, or other, alternative spectral shapes.

\section{Summary and consequences for pulsar surveys}

In \S~\ref{sec:method1} we showed that the relative survey yields of
pulsar surveys at three different frequencies can be explained using
an underlying spectral-index distribution of pulsars which is
Gaussian around a single power law with mean
$\bar{\alpha}=-1.41$ and standard deviation $\bar{\sigma}=0.96$. 
This distribution
is both shallower and wider than the values obtained by both
\citet{lorimer1995} and \citet{maron2000}, although in close agreement
with measurements made by \citet{malofeev2000}, who calculated values
of $\alpha = -1.47, \sigma=0.76$ at observing frequencies of
100-400~MHz. \citet{malofeev2000} explained the difference between
their results and those of \citet{lorimer1995} and \citet{maron2000}
as a steepening of the spectral index at higher radio frequencies, 
however, our results suggest this may in fact be due to incompleteness
and biases in the samples used.

We showed (see Fig.~\ref{fig:combinehists}), perhaps not surprisingly,
that the detected sample of pulsar spectral index is a
strong function of the survey frequency. Surveys carried out at lower
frequencies favour the detection of steeper spectrum pulsars compared
to experiments carried out at higher frequencies. Thus, although we
did not attempt a dynamical study of pulsar evolution, our results
imply that any correlations between spectral index and pulsar
parameters (such as the observation by Lorimer et al.~1995 that
younger pulsars appear to have flatter spectra) needs to be carefully
weighed against the survey frequency they have been selected in.
Our present results imply that the spectral index--age relationship
suggested by Lorimer et al.~(1995)
is most likely a result of observational selection.

In \S~\ref{sec:method2} we investigated the fraction of the pulsar
population that could have spectra which peak at a
frequency of $\sim1~\mathrm{GHz}$.  While the known population of such
sources stands at only five \citep{kijak2011a}, we have shown that a
large number of such sources could exist, though only if the turnover
parameter is very small, $a=-0.5$. Compared to the smallest measured value,
$a=-1.18$ for PSR~B1823$-$13, such a small turnover might not be
noticed for many pulsars unless they are observed at very low
frequencies, for example in the 100~MHz regime.

By allowing the spectral index to take two values, $\alpha_1$ below
1.4~GHz and $\alpha_2$ above, in \S~\ref{sec:method3} we investigated
what fraction of pulsars would display a spectral break. We can
constrain the population of pulsars for which $\alpha_1$ and
$\alpha_2$ differ significantly to be no more than 10\%.

So are we able to place any constraints upon pulsar emission models
from our results? \citet{malofeev1994} and \cite{loehmer2008} used measured spectral
shapes to investigate different models of the pulsar magnetosphere. In so doing, 
they were required to use spectra with a spectral break. We were unable to place
a reasonable constraint upon the values of $\alpha_1$ and $\alpha_2$ in such a model,
and so are unable to rule out such models, or prove them to be valid. However, we have 
shown that only a small fraction of pulsars should deviate from the model of a spectrum 
with a constant power-law slope, and therefore future emission models should be able to 
explain both the size and existence of this sub-population.

\begin{table}
	\begin{center}
	\caption{Observational parameters for the two simulated surveys, the GBNCC survey, and the LOFAR pulsar survey \citep{bws}.}
		\begin{tabular}{lcc}
		\toprule
		& GBNCC & LOFAR \\
		\midrule
		%\multicolumn{2}{c}{Receiver Parameters} \\
		Centre frequency (MHz) & 350 & 143 \\
		Bandwidth (MHz)  & 100 & 47.66 \\
		Channel width (MHz) & 0.0244 & 0.012\\
		Gain ($\mathrm{K~Jy}^{-1}$) & 2.0 & 5.6 \\
		\\
		Integration time (s) & 120 & 1020 \\
		Sampling interval ($\mu$s) & 81.92 & 1310\\
		\\
		Minimum declination (\degree) & $+38$ & $-30$\\
		\bottomrule
		\end{tabular}
		\label{table:simsurveypars}
	\end{center}
\end{table}

\section{Conclusions}

In summary, using simple snapshot models of the pulsar population and
a likelihood analysis we have been able to constrain the underlying
distribution of spectral indices. We find that the survey yields at
three different frequencies can be described by a simple power law
spectrum. The distribution of spectral indices is consistent with
a normal distribution with a mean of --1.4 and unit standard
deviation. Extensions to this simple model were investigated in the
form of Gigahertz-peaked spectrum pulsars and we were able to constrain
the fraction of such sources in the population as a whole to be at
most 10\%. A similar fraction of spectra are consistent with being
double-power-law. Any models of the pulsar emission mechanism should
be able to explain the size and existence of these sub-populations.

Future work to refine this analysis would be to obtain more measurements
of existing pulsar spectra. Only a small fraction of the $\sim2000$ known
radio pulsars have had their flux measured at more than one or two 
observing frequencies. Without more comprehensive flux measurements, it is
not possible to constrain the sizes of any populations with non-standard 
spectral shapes or even to establish the spectral slope.
This would allow a more detailed likelihood
analysis that accounts for the observed distribution of spectra at 
each frequency. One aspect that we did not look at in this work, due to the
simple snapshot approach taken in our simulations, was the case
for any evolution of spectral index with pulsar age. In Lorimer
et al.~(1995), it was proposed that younger pulsars may have flatter
spectra. While our analysis does not directly answer this question, our
demonstration that higher frequency surveys tend to select flatter
spectrum pulsars suggests that any correlation with age is likely
to be weak, since the high-frequency surveys also preferentially
select younger pulsars along the Galactic plane. Further observational 
evidence from large-scale pulsar surveys at several observing frequencies, 
coupled with observational campaigns to accurately measure pulsar spectral 
indices,
will improve our understanding of the spectral index distribution, and
reduce the bias that, from Figure~\ref{fig:combinehists}, seems to exist 
in the current
catalogues.

\section*{ACKNOWLEDGEMENTS}
%The authors would like to thank Tom Hassall and Ryan Lynch for their helpful
%discussions on the LOFAR and GBNCC pulsar surveys, respectively.
JPWV acknowledges financial 
support by the European Research Council for the ERC Starting Grant Beacon under contract 
no. 279702. The authors acknowledge support from WVEPSCoR in the form of a Research 
Challenge Grant. DRL is also supported by the Research Corporation for Scientific 
Advancement as a Cottrell Scholar. We thank the anonymous referee for their helpful 
comments.
\newpage

\bibliography{allrefs}
\bibliographystyle{mnras}

\label{lastpage}

\end{document}